\documentclass[unnumsec,webpdf,contemporary,large]{oup-authoring-template}%

\graphicspath{{Fig/}}

\usepackage{amsmath,amssymb,amsfonts}
\usepackage{longtable}
\usepackage{hyperref}
\usepackage{xurl}
\usepackage{tabularray}
\usepackage{lscape}
 
\theoremstyle{thmstyleone}%
\usepackage{mdframed}
\theoremstyle{thmstyletwo}%
\theoremstyle{thmstylethree}%

\begin{document}

\journaltitle{Bioinformatics}
\DOI{DOI HERE}
\copyrightyear{2023}
\pubyear{2019}
\access{Advance Access Publication Date: Day Month Year}
\appnotes{Paper}

\firstpage{1}


\title{GWAS Summary Statistic Tool: A Meta-Analysis and Parsing Tool for Polygenic Risk Score Calculation}

\author[1,2,$\ast$]{Muhammad Muneeb}
\author[1,2,$\ast$]{David B. Ascher}

\authormark{Muneeb et al.}

\address[1]{\orgdiv{School of Chemistry and Molecular Biology}, \orgname{The University of Queensland}, \orgaddress{\street{Queen Street}, \postcode{4067}, \state{Queensland}, \country{Australia}}}
\address[2]{\orgdiv{Computational Biology and Clinical Informatics}, \orgname{Baker Heart and Diabetes Institute}, \orgaddress{\street{Commercial Road}, \postcode{3004}, \state{Victoria}, \country{Australia}}}

\corresp[$\ast$]{Corresponding authors: David B. Ascher, Email: \href{email:d.ascher@uq.edu.au}{d.ascher@uq.edu.au}; Muhammad Muneeb, Email: \href{email:m.muneeb@uq.edu.au}{m.muneeb@uq.edu.au}}

\received{Date}{0}{Year}
\revised{Date}{0}{Year}
\accepted{Date}{0}{Year}


\abstract{
\textbf{Motivation:} GWAS (genome-wide association study) summary statistic files are essential inputs for polygenic risk score (PRS) calculation, yet identifying suitable files across thousands of catalog entries requires downloading large files and manually inspecting their column structures---a process that is time-consuming and storage-intensive.\\
\textbf{Results:} We present GWASPoker, a phenotype-driven, GWAS-Catalog-specific pre-download triage tool that scans candidate GWAS files for PRS column availability by partial download and header detection, without requiring full-file transfer. Analysing 60,499 records from the GWAS Catalog, 60,281 (99.6\%) contained accessible download links, of which 54,026 (89.6\%) were successfully partially downloaded and parsed across 20 file formats, yielding 724 unique header signatures. Across 13 phenotypes, 84 of 85 manually curated GWAS files (98.8\%) were automatically retrieved and processed. Header validation against fully downloaded files showed exact agreement in 23 of 28 cases (82.1\%).\\
\textbf{Availability and implementation:} GWASPoker is implemented in Python~3 and freely available at \url{https://github.com/MuhammadMuneeb007/GWASPokerforPRS} under the MIT licence. Example outputs and documentation are provided in the repository. The tool was tested on Linux (HPC cluster) with Python~3.8+. The LLM-based code-generation step is entirely optional; a rules-based column-mapping template is provided for fully offline use.
}


\maketitle

\section{Introduction}
GWAS (genome-wide association study) summary statistic files are used to calculate polygenic risk scores (PRS), enabling researchers to make patient risk predictions using association and effect-size information from large datasets without the need to make genotype data publicly available \cite{Uffelmann2021,Mills2019,Witte2010}. These files are publicly available from numerous research groups and offer rich genetic information specific to a phenotype or disease. The procedures, protocols, quality control steps, and methodologies for conducting a GWAS vary depending on the type of phenotype (binary, continuous), and the association results vary depending on the quality, sample size, and the population under consideration \cite{Marees2018,GWAStuto,Turner2011}. This phenomenon results in GWAS summary statistic files with multiple file formats, and the actual information contained in the file can vary, as the purpose of conducting a GWAS can vary, which can be association testing, identifying genetic variants, understanding disease mechanisms, predicting disease risk, personalized medicine, and drug discovery and development. Moreover, GWAS summary statistic files for a specific phenotype can be provided by multiple research groups, and one such collection is the GWAS Catalog, which contains information about more than 60,000 GWAS studies across more than 40,000 diseases and phenotypes \cite{MacArthur2016}. The information in the GWAS Catalog is well-maintained and readily accessible to all. Multiple tools exist for downloading GWAS files from the GWAS Catalog using application programming interface (API) calls, including gwasrapidd \cite{gwasrapidd}, pandasGWAS \cite{Cao2023}, and gwas-download \cite{mikeglou}. Several tools and standards have been developed to support the retrieval, validation, and standardization of GWAS summary statistics. For example, MungeSumstats \cite{Murphy2021MungeSumstats} provides extensive quality control and format standardization for GWAS summary statistics prior to downstream analysis. Similarly, GWASLab \cite{He2023GWASLab} offers a Python toolkit for handling and visualizing GWAS summary statistics. The GWAS Catalog consortium has also proposed a community standard, GWAS-SSF, for representing GWAS summary statistics along with associated metadata \cite{GWASSSF}, and supporting tools, such as gwas-sumstats-tools, provide utilities for reading, formatting, and validating summary statistics files. In addition, alternative data representations, such as GWAS-VCF, enable efficient storage and querying of GWAS summary statistics in indexed formats \cite{Lyon2021GWASVCF}. 

While these tools address downloading, format standardisation, harmonisation, and validation of GWAS summary statistics once they are fully obtained, none provide a phenotype-driven pre-download triage step that checks the readiness of the PRS column via partial file retrieval and header detection. GWASPoker fills this complementary niche: it scans candidate GWAS files from the GWAS Catalog for the presence of PRS-relevant columns before any full-file transfer is performed, enabling rapid selection of suitable datasets without the storage and bandwidth cost of downloading files ranging from 15~MB to 2~GB each. The GWAS files themselves exhibit considerable diversity, encompassing differences in populations studied, sample sizes, genome builds, analysis methods, and reported statistics, which makes automated pre-download triage particularly valuable.


This study developed a tool after analysing 60,499 GWAS summary statistic records sourced from the GWAS Catalog, of which 60,281 (99.6\%) contained accessible download links and 54,026 (89.6\%) were successfully partially downloaded and parsed. The tool first scans the files listed on the GWAS Catalog for a phenotype of interest, retrieves metadata, and partially downloads each candidate file. It then reads each file based on its extension. We included reading functions for 20 file formats. The tool corrects parsing and refines the file for readability, ensuring that the information from the GWAS files is not lost, and identifies column headers in the GWAS file. It also identifies and lists the columns required by PRS tools within the GWAS file. Furthermore, the tool extracts the Digital Object Identifier (DOI) and the article's citation from the PMID, and integrates a Python code generator module to map the original GWAS columns to those required by PRS tools.

\section{Materials and Methods}
\begin{figure*}[!ht]
  \centering
  \includegraphics[width=0.8\textwidth]{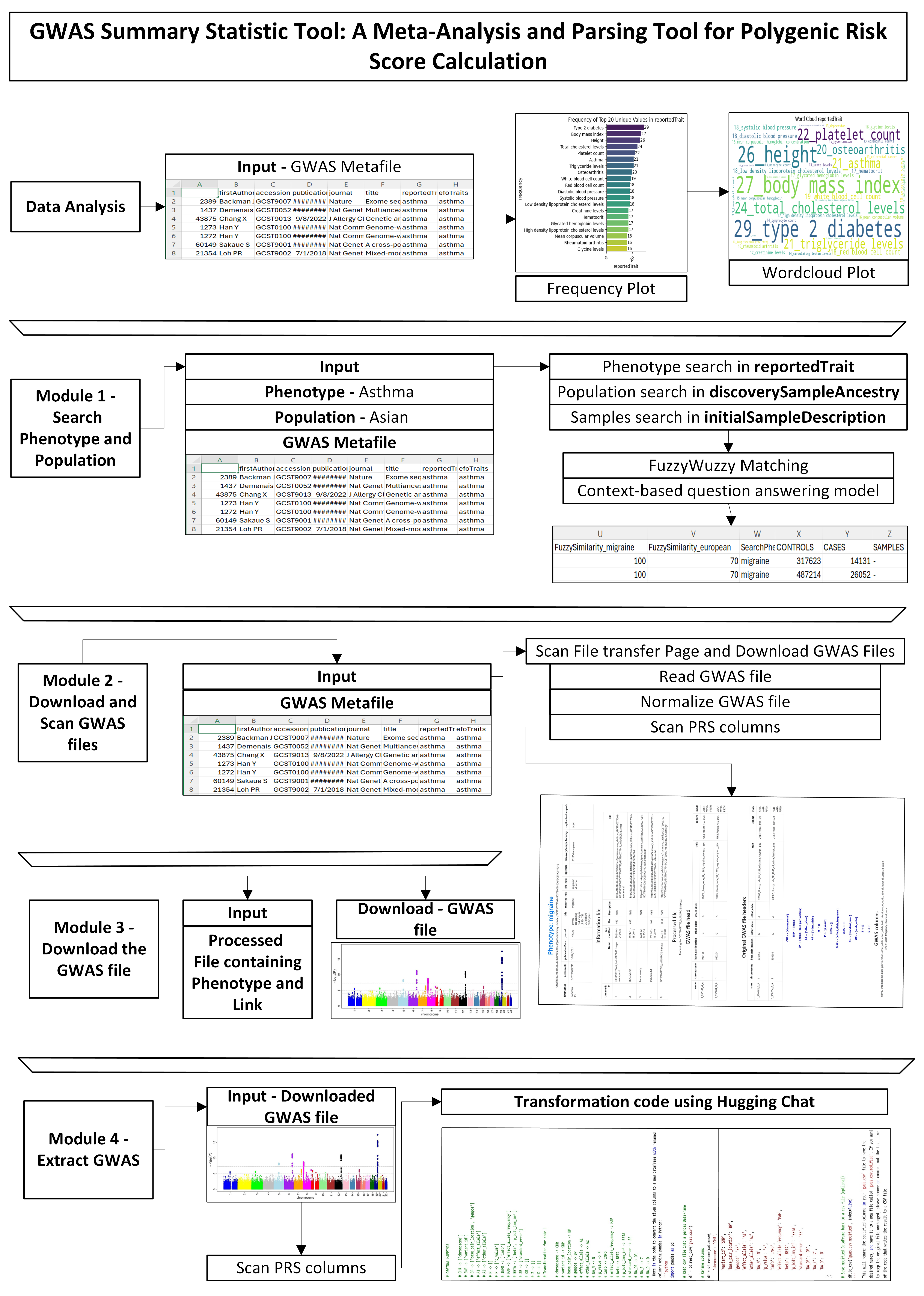}  
  \caption{This figure shows the flowchart of the proposed GWAS downloading and scanning pipeline.}
  \label{flowchart}
\end{figure*}

The GWAS summary statistic metadata was obtained from the GWAS Catalog using the following link: \url{https://www.ebi.ac.uk/gwas/downloads/summary-statistics}, and the file contains information about 60,499 publicly available GWAS files. Figure \ref{flowchart} shows the flowchart of the proposed approach.

\subsection{Module 1 - Search Phenotype and Population}
The GWAS metadata file contains information on 60,499 publicly available GWAS studies at the time of this study. Users can manually process the entire GWAS metadata file to retrieve the necessary information and obtain the corresponding GWAS files for specific phenotypes, or they can search for a specific phenotype in the \texttt{reportedTrait} and \texttt{efoTraits} fields in the GWAS metadata file. Details about a particular population are mentioned in the \texttt{discoverySampleAncestry} field, while information about the number of cases and controls or samples is present in the \texttt{initialSampleDescription} field.

In Module 1, users can specify the phenotype name and a specific population (optional) to retrieve GWAS files that meet those criteria. Fuzzy logic matches the phenotype name in \texttt{reportedTrait} and population name in the \texttt{discoverySampleAncestry} field. Entries with a similarity score of greater than 50 were returned. We included one optional context-based question-answering model \texttt{ahotrod/ electra\_large\_discriminator\_squad2\_512} from Hugging Face that extracts the samples (for continuous phenotypes) and cases and controls (for binary phenotypes) from the \texttt{initialSampleDescription}. The question we considered was the number of cases/controls/samples, and the context was the entry in the \texttt{initialSampleDescription}. This module generates a CSV file for a specific phenotype that contains information about the selected GWAS files, either for the specified phenotype and population or for the phenotype alone.

\subsection{Module 2 - Download and Scan GWAS files}
In Module 2, users manually review the file generated from the previous step and identify GWAS files specific to the phenotypes they intend to download and scan. The same file serves as input for this module. This module partially downloads the GWAS files, removes formatting issues, such as quotes, to ensure readability for Pandas, determines an appropriate delimiter to parse the file either in Pandas or in tabular form, and checks the presence of the columns required by PRS tools in the GWAS file. The module generates an output containing information available on the download page for a specific file, the header of the downloaded GWAS file, the PRS column in the downloaded GWAS file, and details such as the DOI and citation for the GWAS study.

Of 60,499 GWAS summary files, download links were available for 60,281 files, and we partially downloaded and processed them for further analysis.

\subsubsection{Scan File Transfer Page and Download GWAS Files}
The download link for each GWAS file is available in the \texttt{summaryStatistics} field. We parsed the directory listing displayed on the download page into a table, which includes details about the GWAS file and associated files and folders for GWAS analysis. This information comprises GWAS and associated file names, last-modified dates, sizes, descriptions, and URLs. We downloaded this information, and in cases where the actual GWAS file was not directly available on the information page, we searched and downloaded the largest file from the directories listed on the first page (for instance, a sample directory named "harmonised/"). Details regarding the files and folders on the download page, as well as the files contained within these folders, were stored in \texttt{Information.csv} for reference. The \texttt{Information.csv} file contains URLs for all files and folders, facilitating tracking of downloaded GWAS files. We downloaded the largest file among those listed in the "Information" table.
Given that GWAS files range from 15 MB to 2 GB, we used a 10-second timeout for partial downloads to capture file headers and initial rows sufficient for delimiter and column-header detection, while avoiding a full-file transfer.

\subsubsection{Read GWAS file}
Partial downloads sometimes yield incomplete outputs; we therefore implemented format-specific fallback strategies to handle the range of file formats encountered. For each downloaded file, we applied the following parsing approach based on the file extension. For file formats such as .tsv, .txt, .csv, .1\_summary\_table, .ma, .assoc, .meta, .tbl, .linear, and .logistic, we used the cat command to extract information. Subsequently, we read the extracted information using Pandas. If an error occurred during the reading process, the file had to be downloaded manually. If Pandas read the file, we saved the GWAS data.

We adopted a more complicated approach for files in the .gz, .tar, and .zip formats. First, we used the gunzip command to extract the contents of the .gz files and stored them in a temporary CSV file. If this initial attempt failed, we used the cat command as an alternative. If both attempts were unsuccessful, we used a double-gunzip command to perform compression and extraction simultaneously. If all these attempts were unsuccessful, the fallback option was to use the zcat command to extract and save the information to temporary CSV files. We outline the various file formats in which the GWAS files are available in Table \ref{file_extensions}.


\begin{table}[!ht]
\centering
\begin{tabular}{|c|c|}
\hline
\textbf{File Extension} & \textbf{Frequency Count} \\
\hline
.gz & 37833 \\
.tsv & 14859 \\
.txt & 340 \\
.zip & 100 \\
.csv & 3140 \\
.1\_summary\_table & 1 \\
.xlsx & 35 \\
.1 & 3865 \\
.ma & 13 \\
.modified & 2 \\
.assoc & 6 \\
.out & 10 \\
.tar & 2 \\
.GZ & 1 \\
.yaml & 1 \\
.logistic & 2 \\
.RData & 5 \\
.meta & 25 \\
.linear & 3 \\
.tbl & 1 \\
\hline
\end{tabular}
\caption{Frequency count of file extensions of downloaded files. Files in .yaml and.RData formats were discarded from further analysis.}
\label{file_extensions}
\end{table}

\subsubsection{Normalize GWAS file}
The previous step generated a GWAS file for all downloaded and correctly parsed GWAS files.
We removed all occurrences of double quotes from the processed files and replaced tab characters with commas. The next step was to identify the best delimiter for the files from a list of common options, including tabs, commas, semicolons, and spaces, and to select the one that produced more than one column when reading the CSV file with Pandas.

\subsubsection{Scan PRS columns}
We downloaded, extracted, and read 54,026 of 60,281 GWAS files. Among these, we identified 724 unique GWAS header signatures. For PRS calculation, multiple PRS tools require various columns from the GWAS. The \texttt{ldsc munge\_sumstats.py} script \cite{BulikSullivan2015} defines one such set of required columns. We listed 14 columns that should be present in the GWAS to assist with PRS calculation. Although the number of columns and the information in the columns can vary, we considered 14 columns, as shown in Table \ref{prscolumns}. We identified mappings for each column across the 54,026 GWAS files. Once we had the mapping for each column, we checked whether the mapped column name appeared in the GWAS and returned that column name. NA is returned if no matching column is found in the GWAS.

\begin{table}[!ht]
\centering
\begin{tabular}{|l|l|}
\hline
\textbf{Attribute} & \textbf{Number of Unique Values} \\ \hline
Chromosome         & 12                               \\ 
SNP                & 34                               \\ 
Base Pair          & 40                               \\ 
Effect Allele      & 32                               \\ 
Alternative Allele & 34                               \\ 
N                  & 46                               \\ 
P Value            & 219                              \\ 
Info               & 22                               \\ 
MAF                & 109                              \\ 
Beta               & 21                               \\ 
SE                 & 48                               \\ 
OR                 & 16                               \\ 
Z-score            & 17                               \\ 
Direction          & 12                               \\ \hline
\end{tabular}
\caption{The first column shows the column name, and the second column shows the number of unique mappings extracted from the downloaded GWAS files.}
\label{prscolumns}
\end{table}

\subsubsection{Get DOI and Citation}
We converted the PubMed ID (PMID) to a DOI and then submitted it to DOI.org to obtain the citation for a specific PMID.

\subsubsection{Output}
The final output of this module is an information file generated from the original download page, the GWAS header information (which shows the information contained in the downloaded GWAS file), the column mapping information that shows which columns are required for PRS calculation in the GWAS, and the DOI and citation for the PMID. The sample output generated by this step for Asthma is available at the following link: \url{https://github.com/MuhammadMuneeb007/GWASPokerforPRS/blob/main/Results/asthma_output.html}. Researchers can download and view the sample output HTML file.

\subsubsection{Validation of partial-download header detection}
To validate that the partial-download approach correctly identifies GWAS headers, we compared the column headers detected from partially downloaded files with headers obtained from fully downloaded files. We evaluated 30 GWAS datasets; 28 were successfully fully downloaded and assessed, and headers matched exactly in 23 of 28 cases (82.1\% exact header agreement). The five mismatched cases arose from files whose column headers were preceded by comment lines, multi-row metadata blocks, or non-standard character encoding.

\subsection{Module 3 - Download the GWAS file}
After scanning the output file from the previous step for a specific phenotype, the user finalizes the GWAS file to be processed for PRS calculation. This module accepts the phenotype's name and download link for the GWAS file, downloads the file, and saves it in the corresponding directory.

\subsection{Module 4 - Extract GWAS}
This module extracts the GWAS file using the process specified in Module~2 and normalizes the downloaded file by column name, using a rules-based column mapping that associates each detected GWAS column with the corresponding PRS column name, using the synonym dictionaries described in Section~\ref{sec:scancols}. This rules-based mapping requires no internet connection or external service and is the primary output of this module.

As an optional extension, users may pass the generated mapping file to an LLM-based code-generation interface (e.g., the Hugging Face Inference API) to auto-generate Python transformation code. This step is entirely optional, requires no login when using open-access model endpoints, and does not affect the core pipeline. A ready-to-edit mapping template is provided in the repository for users who prefer fully offline operation. Users can open the template for a specific file, make the required modifications, and save the GWAS data in the desired format for PRS calculation.

\section{Module 5 - List PRS Columns}
\label{sec:scancols}
In this module, we provide a script that checks whether the GWAS file contains the columns required by PRS tools and generates a rules-based transformation template. An optional LLM-assisted code-generation step is also available, but is not required for the core pipeline to function.

\section{Results}
We applied the proposed system to analyze 13 phenotypes, and all generated files are available on GitHub. Across the GWAS Catalog metadata file (60,499 records), download links were accessible for 60,281 studies (99.6\%). Using the 10-second timeout partial-download setting, 54,026 of 60,281 files (89.6\%) were successfully partially downloaded and parsed across multiple file formats, and 724 unique GWAS header signatures were identified. Across 13 phenotypes, 85 GWAS files were manually selected, and 84 were successfully retrieved by Module 2 (98.8\% retrieval success). Header validation comparing partially downloaded files with fully downloaded files showed exact header agreement in 23 of 28 evaluated GWAS datasets (82.1\%).
Table~\ref{tab:evaluation} summarises the quantitative evaluation metrics across all experiments.

\begin{table*}[!ht]
\centering
\begin{tabular}{|l|c|}
\hline
\textbf{Evaluation metric} & \textbf{Result} \\
\hline
Accessible download links & 60,281 / 60,499 (99.6\%) \\ \hline
Successfully partially downloaded and parsed & 54,026 / 60,281 (89.6\%) \\ \hline
Phenotype-based retrieval success (13 phenotypes) & 84 / 85 (98.8\%) \\ \hline
Exact header agreement (partial vs full download) & 23 / 28 (82.1\%) \\
\hline
\end{tabular}
\caption{Quantitative evaluation of GWASPoker across GWAS Catalog records, phenotype-based retrieval experiments, and header validation tests.}
\label{tab:evaluation}
\end{table*}

Table \ref{tab:gwas_counts} shows the number of GWAS results returned by Module 1 when searching for a specific phenotype in the GWAS metafile. The second column indicates the number of remaining GWAS files after manual filtering, before they are passed to Module 2. We filtered out GWAS that did not precisely match the phenotype name, those containing gene-burden information, those specific to particular body parts, and those with repeated population studies. We selected at least one GWAS from each population with the maximum number of discovery samples for inclusion in the final analysis. The last column shows the files retrieved using the proposed methodology, indicating the number of files retrieved from Module 2.


\begin{table*}[!ht]
\centering
\begin{tabular}{|l|c|c|c|}
\hline
\textbf{Directory} & \textbf{GWAS metafile} & \textbf{Manually listed} & \textbf{Retrieved files} \\ \hline
asthma & 43 & 10 & 10 \\ \hline
blood\_pressure\_medication & 293 & 2 & 2 \\ \hline
body\_mass\_index\_bmi & 55 & 14 & 14 \\ \hline
cholesterol\_lowering\_medication & 360 & 1 & 1 \\ \hline
depression & 59 & 12 & 12 \\ \hline
gastro\_oesophageal\_reflux\_gord\_gastric\_reflux & 14 & 8 & 8 \\ \hline
hayfever\_allergic\_rhinitis & 37 & 2 & 2 \\ \hline
high\_cholesterol & 333 & 5 & 5 \\ \hline
hypertension & 55 & 11 & 11 \\ \hline
hypothyroidism\_myxoedema & 28 & 2 & 2 \\ \hline
irritable\_bowel\_syndrome & 61 & 4 & 4 \\ \hline
migraine & 28 & 4 & 4 \\ \hline
osteoarthritis & 107 & 10 & 9 \\ \hline
\end{tabular}
\caption{The GWAS metafile column indicates the total number of GWAS retrieved from the GWAS metafile (Module 1), while Manually listed shows the number of files manually selected, and Retrieved files shows the file retrieved using Module 2.}
\label{tab:gwas_counts}
\end{table*}

\section{Conclusion}
In this manuscript, we propose a set of code that enables researchers to search and scan the GWAS Catalog without downloading GWAS files. The pipeline lists which PRS columns are available in the GWAS file, and the user can scan multiple GWAS files to determine which file to download and use for further processing. We executed the pipeline on 13 different phenotypes and performed the proposed analysis. The pipeline completed analysis for each phenotype in 4 to 5 hours, considerably faster than manual file downloading and column inspection.

\section{Availability and implementation}
GWASPoker is implemented in Python~3 and is freely available at \url{https://github.com/MuhammadMuneeb007/GWASPokerforPRS} under the MIT licence. Full documentation, example output files for 13 phenotypes, and a rules-based column-mapping template for offline use are provided in the repository. The tool was developed and tested on a Linux HPC environment with Python~3.8+. The LLM-based code-generation step (Modules~4 and~5) is entirely optional and does not affect the core partial-download and header-detection functionality.

\section{Competing interests}
The authors declare that they have no competing interests.

\section{Author contributions statement}
M.M. wrote the first draft of the manuscript and wrote and tested the code. M.M. analyzed the results. D.A. reviewed and edited the manuscript. All authors contributed to the manuscript's methodology.

\section{Data availability}
The dataset used in this study is available from the GWAS Catalog (\url{https://www.ebi.ac.uk/gwas/}), and the code is available on GitHub (\url{https://github.com/MuhammadMuneeb007/GWASPokerforPRS}). The results and the output files are available on the GitHub repository \url{https://github.com/MuhammadMuneeb007/GWASPokerforPRS/tree/main/Results}.

\section{Acknowledgments}
Not applicable.
\bibliographystyle{unsrt}
\bibliography{reference}

@article{Mills2019,
  title = {A scientometric review of genome-wide association studies},
  volume = {2},
  ISSN = {2399-3642},
  url = {http://dx.doi.org/10.1038/s42003-018-0261-x},
  DOI = {10.1038/s42003-018-0261-x},
  number = {1},
  journal = {Communications Biology},
  publisher = {Springer Science and Business Media LLC},
  author = {Mills,  Melinda C. and Rahal,  Charles},
  year = {2019},
  month = jan 
}

@article{MacArthur2016,
  title = {The new NHGRI-EBI Catalog of published genome-wide association studies (GWAS Catalog)},
  volume = {45},
  ISSN = {1362-4962},
  url = {http://dx.doi.org/10.1093/nar/gkw1133},
  DOI = {10.1093/nar/gkw1133},
  number = {D1},
  journal = {Nucleic Acids Research},
  publisher = {Oxford University Press (OUP)},
  author = {MacArthur,  Jacqueline and Bowler,  Emily and Cerezo,  Maria and Gil,  Laurent and Hall,  Peggy and Hastings,  Emma and Junkins,  Heather and McMahon,  Aoife and Milano,  Annalisa and Morales,  Joannella and Pendlington,  Zoe May and Welter,  Danielle and Burdett,  Tony and Hindorff,  Lucia and Flicek,  Paul and Cunningham,  Fiona and Parkinson,  Helen},
  year = {2016},
  month = nov,
  pages = {D896–D901}
}

@article{Cao2023,
  title = {pandasGWAS: a Python package for easy retrieval of GWAS catalog data},
  volume = {24},
  ISSN = {1471-2164},
  url = {http://dx.doi.org/10.1186/s12864-023-09340-2},
  DOI = {10.1186/s12864-023-09340-2},
  number = {1},
  journal = {BMC Genomics},
  publisher = {Springer Science and Business Media LLC},
  author = {Cao,  Tianze and Li,  Anshui and Huang,  Yuexia},
  year = {2023},
  month = may 
}

@article{Turner2011,
  title = {Quality Control Procedures for Genome‐Wide Association Studies},
  volume = {68},
  ISSN = {1934-8258},
  url = {http://dx.doi.org/10.1002/0471142905.hg0119s68},
  DOI = {10.1002/0471142905.hg0119s68},
  number = {1},
  journal = {Current Protocols in Human Genetics},
  publisher = {Wiley},
  author = {Turner,  Stephen and Armstrong,  Loren L. and Bradford,  Yuki and Carlson,  Christopher S. and Crawford,  Dana C. and Crenshaw,  Andrew T. and de Andrade,  Mariza and Doheny,  Kimberly F. and Haines,  Jonathan L. and Hayes,  Geoffrey and Jarvik,  Gail and Jiang,  Lan and Kullo,  Iftikhar J. and Li,  Rongling and Ling,  Hua and Manolio,  Teri A. and Matsumoto,  Martha and McCarty,  Catherine A. and McDavid,  Andrew N. and Mirel,  Daniel B. and Paschall,  Justin E. and Pugh,  Elizabeth W. and Rasmussen,  Luke V. and Wilke,  Russell A. and Zuvich,  Rebecca L. and Ritchie,  Marylyn D.},
  year = {2011},
  month = jan 
}

@article{BulikSullivan2015,
  title = {Relationship between LD Score and Haseman-Elston Regression},
  url = {http://dx.doi.org/10.1101/018283},
  DOI = {10.1101/018283},
  publisher = {Cold Spring Harbor Laboratory},
  author = {Bulik-Sullivan,  Brendan},
  year = {2015}, 
 journal ={biorxiv}
}

@article{Witte2010,
  title = {Genome-Wide Association Studies and Beyond},
  volume = {31},
  ISSN = {1545-2093},
  url = {http://dx.doi.org/10.1146/annurev.publhealth.012809.103723},
  DOI = {10.1146/annurev.publhealth.012809.103723},
  number = {1},
  journal = {Annual Review of Public Health},
  publisher = {Annual Reviews},
  author = {Witte,  John S.},
  year = {2010},
  month = mar,
  pages = {9–20}
}

@article{He2023GWASLab,
  title   = {{GWASLab}: a Python package for processing and visualizing {GWAS} summary statistics},
  author  = {He, Yunye and Koido, Masaru and Shimmori, Yuka and Kamatani, Yoichiro},
  journal = {Jxiv},
  year    = {2023},
  doi     = {10.51094/jxiv.370}
}

@article{Lyon2021GWASVCF,
  title   = {The variant call format provides efficient and robust storage of {GWAS} summary statistics},
  author  = {Lyon, Matthew S and Andrews, Stephen J and Elsworth, Benjamin and Gaunt, Tom R and Hemani, Gibran and Marcora, Elisa},
  journal = {Genome Biology},
  year    = {2021},
  doi     = {10.1186/s13059-020-02248-0}
}

@article{Murphy2021MungeSumstats,
  title   = {MungeSumstats: a {Bioconductor} package for the standardization and quality control of many {GWAS} summary statistics},
  author  = {Murphy, Alan E and Schilder, Brian M and Skene, Nathan G},
  journal = {Bioinformatics},
  year    = {2021},
  doi     = {10.1093/bioinformatics/btab665}
}

@article{Marees2018,
  doi = {10.1002/mpr.1608},
  url = {https://doi.org/10.1002/mpr.1608},
  year = {2018},
  month = feb,
  publisher = {Wiley},
  volume = {27},
  number = {2},
  pages = {e1608},
  author = {Andries T. Marees and Hilde de Kluiver and Sven Stringer and Florence Vorspan and Emmanuel Curis and Cynthia Marie-Claire and Eske M. Derks},
  title = {A tutorial on conducting genome-wide association studies: Quality control and statistical analysis},
  journal = {International Journal of Methods in Psychiatric Research}
}

@article{Uffelmann2021,
  doi = {10.1038/s43586-021-00056-9},
  url = {https://doi.org/10.1038/s43586-021-00056-9},
  year = {2021},
  month = aug,
  publisher = {Springer Science and Business Media {LLC}},
  volume = {1},
  number = {1},
  author = {Emil Uffelmann and Qin Qin Huang and Nchangwi Syntia Munung and Jantina de Vries and Yukinori Okada and Alicia R. Martin and Hilary C. Martin and Tuuli Lappalainen and Danielle Posthuma},
  title = {Genome-wide association studies},
  journal = {Nature Reviews Methods Primers}
}

\end{document}